# Does the Redshift Distribution of *Swift* Long GRBs Trace the Star-Formation Rate?

Ali M. Hasan, Walid J. Azzam

Department of Physics, College of Science, University of Bahrain, Sakhir, Bahrain
Email: wjazzam@uob.edu.bh, wjazzam@gmail.com

## Abstract

Gamma-ray bursts (GRBs) are extremely powerful explosions that have been traditionally classified into two categories: long bursts (LGRBs) with an observed duration $T_{90} > 2$ s, and short bursts (SGRBs) with an observed duration $T_{90} < 2$ s, where $T_{90}$ is the time interval during which 90% of the fluence is detected. LGRBs are believed to emanate from the core-collapse of massive stars, while SGRBs are believed to result from the merging of two compact objects, like two neutron stars. Because LGRBs are produced by the violent death of massive stars, we expect that their redshift distribution should trace the star-formation rate (SFR). The purpose of our study is to investigate the extent to which the redshift distribution of LGRBs follows and reflects the SFR. We use a sample of 370 LGRBs taken from the *Swift* catalog, and we investigate different models for the LGRB redshift distribution. We also carry out Monte Carlo simulations to check the consistency of our results. Our results indicate that the SFR can describe the LGRB redshift distribution well for high redshift bursts, but it needs an evolution term to fit the distribution well at low redshift.

## 1. Introduction

Although more than 50 years have passed since the first paper on gamma-ray bursts (GRBs) was published [1], we still have many questions regarding their nature and characteristics. They are the focus of many research studies due to their potential as cosmological probes [2]-[8]. GRBs are traditionally classified into two main types: long GRBs (LGRBs) with $T_{90} > 2$ s and short GRBs (SGRBs) with $T_{90} < 2$ s [9], where $T_{90}$ is the observed duration during which 90% of the fluence is detected. It is theorized that the origin of LGRBs is the core-collapse of massive stars, while SGRBs are thought to originate from compact object mergers [10] [11] [12]. Hence, it is expected that there will be a correlation between the star formation rate (SFR) and the LGRB formation rate, which encouraged researchers to investigate the redshift distribution of GRBs, and especially that of LGRBs.

However, it is important to note that the connection between the SFR and the LGRB redshift distribution is a controversial one, with many conflicting results in the literature. The controversy lies in whether there is an observed excess of GRBs at low or high redshift compared to what is implied by the SFR. The study by [13] is one of the first major contributions to this problem. Although the connection between the SFR and LGRB redshift distribution had been studied before, see for example [14], the study by [13] is one of the first to account for completeness and to place constraints on the GRB luminosity function and its evolution. Their study found evidence for a possible luminosity or density evolution of LGRBs compared to what is expected from the SFR [13]. A recent study, but with a larger data sample, by [15] found similar results. An earlier study by [16], which utilized 127 LGRBs observed by the *Swift* satellite, found an excess in GRBs at low redshift compared to the SFR. Although some studies backed the low-redshift excess claims [17] [18] [19] [20], other studies found an excess in GRBs, not at low redshift, but at high redshift [21] [22].

Moreover, many studies have opposed the excess GRB accounts, especially at low redshifts. The study by [23] was one of the earliest to do so, where they carried out Monte Carlo simulations and compared their results with previously published claims of low redshift excess – the aforementioned [16] [19] studies. They showed that the excess GRBs result from the incompleteness of the dataset, and that a complete dataset shows no excess in GRBs. The study by [23] also conducted a non-parametric $C^-$ investigation. However, their non-parametric study seems misleading as discussed by [24] who made an in-depth study about the non-parametric methods and how their misuse can lead to incorrect results. They analyzed datasets of previous GRB luminosity/redshift evolution studies, like the previously mentioned [16] [18] [20] [23]. They found that many studies underestimated the detection threshold of GRB satellites which resulted in incomplete datasets, and hence showed an excess in GRBs. They also created a Monte Carlo simulation, which they found to agree with the Monte Carlo simulations conducted by [23]; in other words, excess GRBs resulted from incomplete datasets. The investigation by [25] also found no excess GRBs at any redshift.

More recently, the excess GRBs at low redshift claims have started to reemerge. They are mainly driven by studies conducted by [26] [27]. In the first study, [26] formed two sets of GRBs composed of 54 and 79 LGRBs, with 61% and 67% completeness, respectively. They found an excess of LGRBs at low redshift, even with the completeness of the dataset taken into consideration. The second study [27] focused on potential reasons for this excess of LGRBs. They split their datasets into low and high luminosity LGRBs. They found that the high luminosity LGRB dataset follows the SFR without any excess, while the low luminosity LGRB dataset did show an excess of LGRBs at low redshifts. They argue that their results may hint at new origins for low luminosity LGRBs, thus disconnecting the LGRB formation rate from the SFR. Interestingly, there were some recent reports that suggest that LGRBs can be produced by compact object mergers, instead of massive star collapse [28] [29]. Moreover, there have been studies that investigated the effect of metallicity on the LGRB distribution and have found it to be influential [30] [31]. Perhaps these can be potential reasons for the excess GRB dilemma. Although as it stands, there is no clear answer to this problem.

The aim of our current study is to investigate the distribution of LGRBs with redshift and to fit it with several proposed models. A full LGRB data set is used without constraints, including biases in detection, to study how it is linked to the star formation rate. We aim to examine the nature of the evolution term and see how it impacts the redshift distribution of LGRBs. Section 2 provides a detailed description of our method, while Section 3 presents our results and analysis. We provide a brief conclusion in the last section.

## 2. Method

The data sample that we use is collected from the *Swift* catalog[1]. The data sample is composed of 370 LGRBs with $T_{90} > 2$ s. The dataset is found in Table A1 in Appendix A. We divided the sample into 30 bins for $0 \leq z \leq 10$ with equal bin sizes $dz = \frac{1}{3}$, and we then normalized the distribution. Mathematically, the distribution of GRBs can be written as:

$$\Phi(z) = \frac{dN}{dz} = \frac{dN}{dV}\frac{dV}{dz} = A\frac{\psi_*(z)}{1+z}\phi(z)\frac{dV(z)}{dz} \qquad (1)$$

where $\psi_*(z)$ is the star-formation rate density (SFRD), $\phi(z)$ is a term that contains all contributions to the distribution other than the SFRD, $\frac{dV}{dz}$ is the co-moving volume element and $A$ is a normalization constant. In this paper, we do not assign any specific "meaning" to $\phi(z)$ and $A$. All possible contributions such as the luminosity function, density evolution, sensitivity of the detectors and others are included in $\phi(z)$. Our aim is to gain insight on the nature of $\phi(z)$ without presuming any prior connections or contributions from other sources. This is done to focus our work on the link to the SFRD, and to evaluate how significant the other contributions are to the redshift distribution of LGRBs. We caution here that since detection bias is also considered in Equation (1), then the equation represents the GRB detection distribution, rather than the formation distribution.

The SFRD that is typically used in the literature in units of $M_\odot \cdot yr^{-1} \cdot Mpc^{-3}$

---
[1]https://swift.gsfc.nasa.gov/archive/grb_table/.

is [32] [33]:

$$\psi_*(z) = \frac{0.0157 + 0.118z}{1 + \left(\frac{z}{3.23}\right)^{4.66}} \quad (2)$$

We adopt the ΛCMD model in this study, which gives:

$$\frac{dV}{dz} = \frac{4\pi c D_L^2}{H_0 (1+z)^2 \sqrt{\Omega_M (1+z)^3 + \Omega_\Lambda}} \quad (3)$$

where $D_L$ is the luminosity distance, which is given by:

$$D_L(z) = \frac{(1+z)c}{H_0} \int_0^z \frac{dz}{\sqrt{\Omega_M (1+z)^3 + \Omega_\Lambda}} \quad (4)$$

and $H_0 = 70.8$ km/s/MPc, $\Omega_M = 0.27$, and $\Omega_\Lambda = 1 - \Omega_M = 0.73$ are the adopted cosmological parameters.

We investigated and fitted our data sample with multiple models and the goodness of the fit for each model was determined using the least $\chi^2$ test:

$$\chi^2 = \sum_i \frac{\left(N_i^{\text{swift}} - N_i^{\text{cal}}\right)^2}{\sigma_i^2} \quad (5)$$

where $\sigma_i = \sqrt{N_i^{\text{swift}}}$ is the Poisson error (68% confidence). The best fitting model is the one that gave the least $\chi^2$ value.

The first model that we tried was a "no contributions" model, $\phi(z) = $ constant, with the SFRD and the cosmological model being the only contributors. Then, we tested five different contribution terms $\phi(z)$: a power law, a broken power law, a triple power law, an exponential term, and an exponential-power law, as follows:

$$\phi_{\text{pow}}(z) = (1+z)^\alpha \quad (6)$$

$$\phi_{\text{broken}}(z) = \begin{cases} \dfrac{(1+z)^\alpha}{(1+r_1)^\alpha} & \text{for } z \leq r_1 \\ \dfrac{(1+z)^\beta}{(1+r_1)^\beta} & \text{for } z > r_1 \end{cases} \quad (7)$$

$$\phi_{\text{triple}}(z) = \begin{cases} \dfrac{(1+z)^\alpha}{(1+r_1)^\alpha} & \text{for } z \leq r_1 \\ \dfrac{(1+z)^\beta}{(1+r_1)^\beta} & \text{for } r_1 < z \leq r_2 \\ \dfrac{(1+r_2)^\beta}{(1+r_1)^{\delta_2}} \dfrac{(1+z)^\gamma}{(1+r_2)^\gamma} & \text{for } r_2 \leq z \end{cases} \quad (8)$$

$$\phi_{\text{exp}}(z) = e^{\mu z} \quad (9)$$

$$\phi_{\text{exp-pow}}(z) = (1+z)^\alpha e^{\mu z} \quad (10)$$

where $\alpha, \beta, \gamma, \mu, r_1$, and $r_2$ are the optimization parameters[2]. Although we do not force any specific form on $\phi(z)$, the expressions that we selected and chose to investigate are based on our intuitive expectations of the insight that they might provide. The broken and triple power laws were selected to give insight on how the distribution behaves at different redshift ranges. The power and exponential laws were selected to be compared with the broken/triple power law to see if the evolution is region specific, or generic to the whole distribution.

In this work, we chose two optimization methods. The first was a fitting method. The fitting was done for $\alpha, \beta, \gamma, \mu$, while $r_1$ and $r_2$ were found by looping over them and finding the optimal values that minimize $\chi^2$. The second method involved using the Metropolis-Hastings Markov Chain Monte Carlo (MCMC) algorithm [34]. The method used is described by Foreman-Mackey and Hogg and utilizes their Python libraries emcee[3] and corner.py[4] [35] [36]. The log of the likelihood function that we wish to maximize is of the form:

$$\ln \mathcal{L} = -\frac{1}{2} \sum_i \left[ \frac{\left(N_i^{\text{swift}} - N_i^{\text{cal}}\right)^2}{\sigma_i^2} \right] = -\frac{1}{2} \chi^2. \tag{11}$$

Then $\chi^2$ can be calculated as follows:

$$\chi^2 = -2 \ln \mathcal{L} \tag{12}$$

Here maximizing the likelihood function is equivalent to minimizing the $\chi^2$ value. However, since the models used include different numbers of parameters, the Akaike information criterion (AIC) was used to assess the goodness of the fits and to compare the different models [37] [38]:

$$\text{AIC} = 2k + \chi^2 = 2k - 2\ln(\mathcal{L}) \tag{13}$$

where $k$ is the number of parameters. The better fitting model is the one that minimizes the AIC value.

We then used the MCMC simulations to fit the data to the following:

1) A new SFRD-like function—call it the GRB detection density function (GRB-DD):

$$\psi_{\text{GRB}}(z) = \frac{\alpha + \beta z}{1 + \left(\frac{z}{\gamma}\right)^\delta} \tag{14}$$

2) A broken power law (see Equation (7)).
3) A triple power law (see Equation (8)).
4) An exponential-power law (see Equation (10)).

For each case, we report the mean of the probability distribution and its deviation, the best fitting value, the $\chi^2$ value, and the AIC.

## 3. Results

The normalized redshift distribution of the data sample used is shown in **Figure 1(a)**.

---

[2]Note that the best fitting parameters for different cases are expected to be different.
[3]https://emcee.readthedocs.io/en/stable/.
[4]https://corner.readthedocs.io/en/latest/.

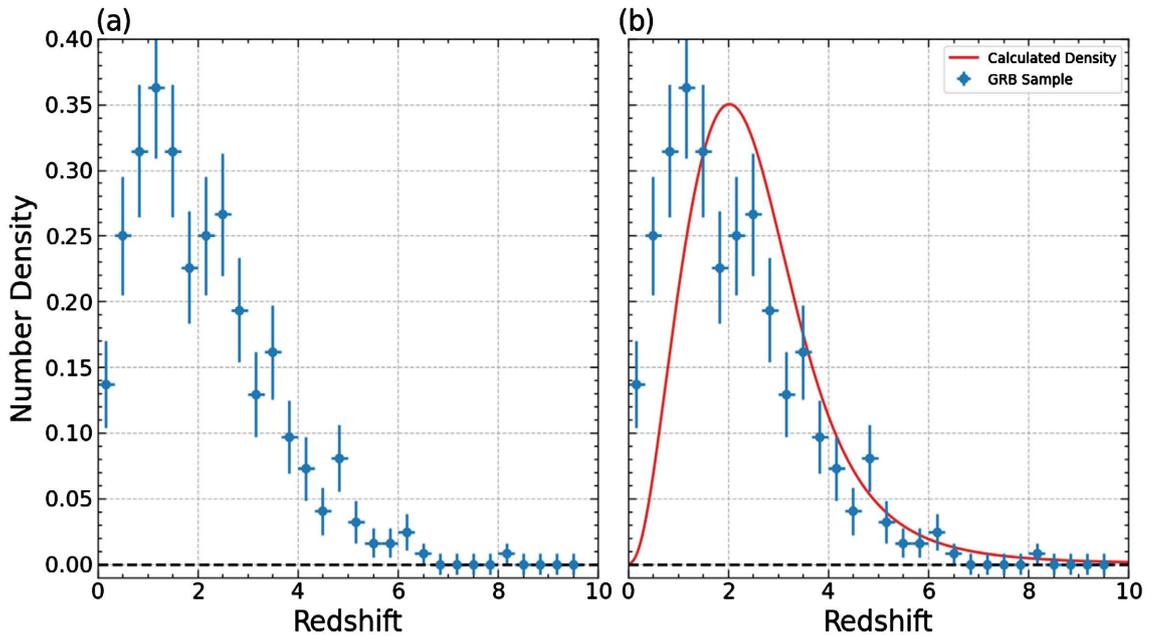

**Figure 1.** Shows (a) the redshift distribution of the GRB data set and (b) a comparison between the calculated no evolution GRB number density and the redshift distribution. Note that both the calculated density and the redshift distribution were normalized.

The main highlight of the results is the immediate spike of the number density, peaking at $z \sim 1.25$. Comparing that with the density calculated with no evolution (Figure 1(b)), we can see that the "no evolution" density peaks at $z = 2$ with a subtle increase compared to the *Swift* data. Focusing on the low and high redshift regions, we notice the calculated density matches the *Swift* distribution at high redshifts. However, at low redshifts it is a shifted version of the *Swift* data, shifted to the right by about $\Delta z \sim 0.5$. To solve this issue, one might shift the density left but that would not solve the problem, as then it would underestimate the number density at $z \gtrsim 4$, which presents the main issue with this model. We need a term $\phi(z)$ to be included in the density that scales it at low redshift while maintaining a good fit at high redshifts, like the SFRD (Figure 1(b)).

### 3.1. Fitting Results

We fitted the data sample with the proposed functions (Equations (6) to (10)), and we extracted the relevant parameters. The results are shown in Table 1. Based on the AIC test, the best fit occurred when we used the triple power law, the second best fit when we used the broken power law, and the third best fit when we used the exponential-power law. The power law and the exponential cases show very poor fitting. These results highlight the need for an additional $\phi(z)$ factor and suggest that it is redshift "region" specific.

### 3.2. MCMC Results

The MCMC results are shown in Table 2. It is notable that the results agree mostly with the fitting results found (within $1\sigma$) and lead to the same conclu-sion. Thus, we can conclude that the results are consistent with each other. All proposed models show good fitting, with the triple power law showing the best fit. Figure 2 plots all the models for comparison.

Now focusing on the variation between the fitting of the contributions at low and high redshifts, we notice that the results show a significant difference between the two redshift regions. This is clearly exhibited by the fitting and MCMC results, particularly in the broken power law and the triple power law cases. These cases mainly split the distribution into two regions, separated at around $z \sim 1.75$. Below $z \lesssim 1.75$, there is a high contribution from $\phi(z)$ with powers $\alpha = -2.56^{+0.41}_{-0.54}$ (broken power law), and $\alpha = -5.36^{+2.60}_{-2.12}$ and $\beta = -2.01^{+0.91}_{-0.50}$ (triple power law). Meanwhile, for $z \gtrsim 1.75$, the contribution from $\phi(z)$ becomes much less impactful with values $\beta = 0.00^{+0.30}_{-0.31}$ (broken power law) and $\gamma = 0.06^{+0.30}_{-0.29}$ (triple power law).

We see the same behavior with the GRB-DD fit. The values for $\beta, \gamma$, and $\delta$ that we found coincide with the literature reported values (see Equation (2)) within $1\sigma$ ($2\sigma$ in the case of $\beta$). The major difference is in $\alpha$, with the value found by MCMC being $\alpha = 2.88^{+1.28}_{-1.37}$ compared to the literature reported value of $\alpha = 0.0157$ (see Equation (2)). Figure 3 shows a plot of the GRB-DD and SFRD for reference. We see that the difference between the two functions is very large at low redshifts but becomes smaller as the redshift increases to $z > 4$.

**Table 1.** The parameters calculated by fitting for the different cases studied with their standard deviations (if applicable).

| | $\alpha$ | $\beta$ | $\gamma$ | $\mu$ | $r_1$ | $r_2$ | $\chi^2$ | AIC |
|---|---|---|---|---|---|---|---|---|
| Only SFRD Contribution | - | - | - | - | - | - | 83.948 | --- |
| Power Law | $-1.242 \pm 0.166$ | - | - | - | - | - | 50.964 | 52.964 |
| Broken Power Law | $-2.845 \pm 0.276$ | $0.043 \pm 0.244$ | - | - | 1.780 | - | 15.748 | 21.748 |
| Triple Power Law | $-7.117 \pm 1.078$ | $-2.595 \pm 0.297$ | $-0.016 \pm 0.249$ | - | 0.500 | 1.800 | 8.184 | 18.184 |
| Exponential | - | - | - | $-0.271 \pm 0.054$ | - | - | 66.960 | 68.960 |
| Exponential-Power | $-3.742 \pm 0.466$ | - | - | $0.767 \pm 0.134$ | - | - | 23.684 | 27.684 |

**Table 2.** The MCMC optimization results. The table includes the mean of the probability distribution and its deviation, the best fitting value, the $\chi^2$ value, and the AIC value for each case. The corresponding contour plots of each case can be found in Appendix B.

| | | $\alpha$ | $\beta$ | $\gamma$ | $\delta$ | $\mu$ | $r_1$ | $r_2$ | $\chi^2$ | AIC |
|---|---|---|---|---|---|---|---|---|---|---|
| GRB-DD Contribution | Most Probable | $2.88^{+1.28}_{-1.37}$ | $-0.20^{+0.16}_{-0.13}$ | $4.74^{+0.69}_{-0.60}$ | $4.94^{+1.26}_{-1.23}$ | - | - | - | 17.499 | 25.499 |
| | Best Fit | 3.190 | $-0.365$ | 5.434 | 5.463 | - | - | - | | |
| Broken Power Law | Most Probable | $-2.56^{+0.41}_{-0.54}$ | $0.00^{+0.30}_{-0.31}$ | - | - | - | $1.71^{+0.23}_{-0.30}$ | - | 17.145 | 23.145 |
| | Best Fit | $-2.485$ | 0.082 | - | - | - | 1.767 | - | | |
| Triple Power Law | Most Probable | $-5.36^{+2.60}_{-2.12}$ | $-2.01^{+0.91}_{-0.50}$ | $0.06^{+0.30}_{-0.29}$ | - | - | $0.55^{+0.73}_{-0.19}$ | $1.89^{+0.43}_{-0.22}$ | 10.657 | 20.657 |
| | Best Fit | $-6.280$ | $-2.111$ | 0.052 | - | - | 0.520 | 1.855 | | |
| Exponential-Power | Most Probable | $-3.23^{+0.48}_{-0.44}$ | - | - | - | $0.67^{+0.12}_{-0.13}$ | - | - | 26.523 | 30.523 |
| | Best Fit | $-3.291$ | - | - | - | 0.686 | - | - | | |

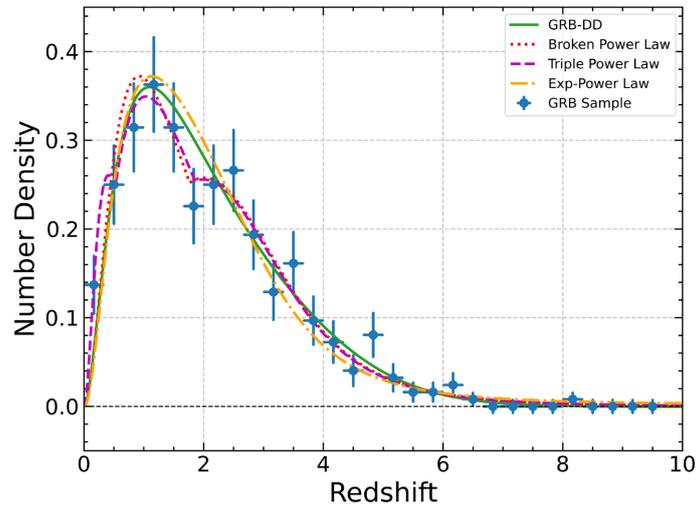

**Figure 2.** Shows number density vs redshift for all the models fit with MCMC. The best fit values were chosen for the plot of each density. It is noticeable that all the models meet at high redshifts, and their differences only occur at low redshifts.

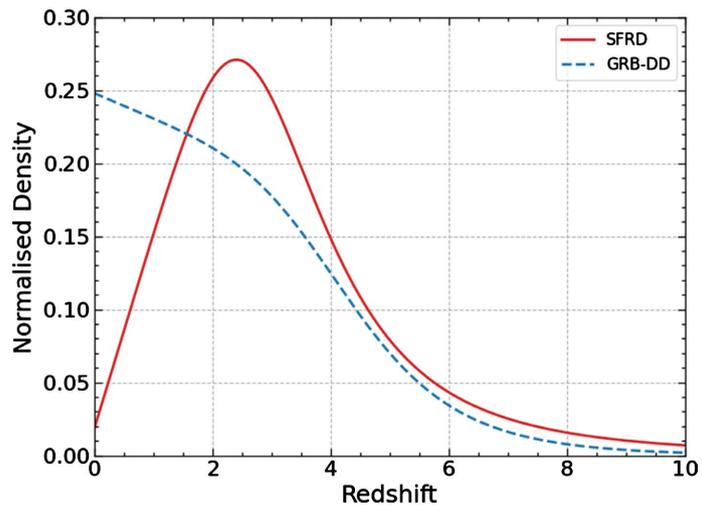

**Figure 3.** The SFRD compared to the calculated GRB-DD.

To summarize, our results show that the *Swift* data are consistent with the SFRD at high redshifts, and only requires an evolution term that enhances it at low redshifts. This means that if the LGRBs are to follow the SFR, then whatever factors that might affect their detection must have a stronger influence at low redshift than at high redshift. Also, it is notable that high redshift LGRBs detected by *Swift* are predominantly high luminosity ones. Hence, since our results follow the SFR at high redshifts, they support the findings of [27].

One important factor to mention that may affect the results is the classification of GRBs. In our analysis, we followed the traditional definition of LGRBs ($T_{90} > 2$ s). This definition is widely debated among researchers with many of the newer studies pushing for updated classifications [39]-[45]. Thus, in the future one might investigate this issue further when studying the connection between LGRBs and SFR, by considering a different classification and seeing if it would change the results found in the literature.

## 4. Conclusion

In this paper, different models for the LGRB redshift distribution were studied. A complete data sample of *Swift* long GRBs with known redshifts was used without imposing any constraints. Our study shows that the SFRD can describe the LGRB redshift distribution for high redshift but needs an evolution term, to fit the distribution well, at low redshift. Although the data selected may have detection biases, primarily at low redshift, our results support the connection between long GRBs and stellar core-collapse. Our results indicate that the redshift distribution of the data sample fits the SFRD well at high redshifts. However, the issue of excess GRBs at low redshifts persists, because as far as our results go, they do not give us any indication about the nature of the extra contribution term $\phi(z)$. To verify our results, future studies should consider data sets from other satellites or a combined data sample from more than one satellite with a focus on high redshift GRBs.

## Conflicts of Interest

The authors declare no conflicts of interest regarding the publication of this paper.

# Appendix A. GRB Dataset

Table A1. Shows the GRB dataset used.

| GRB | $T_{90}$ | Redshift |
|---|---|---|
| 230506C | 31.00 | 3.7 |
| 230414B | 25.98 | 3.568 |
| 230116D | 41.00 | 3.81 |
| 221226B | 3.44 | 2.694 |
| 221110A | 8.98 | 4.06 |
| 221009A | 1068.40 | 0.1505 |
| 220611A | 57.00 | 2.3608 |
| 220521A | 13.55 | 5.6 |
| 220117A | 49.81 | 4.961 |
| 220101A | 173.36 | 4.61 |
| 211207A | 3.73 | 2.272 |
| 211024B | 603.5 | 1.1137 |
| 10822A | 180.8 | 1.736 |
| 210731A | 22.51 | 1.2525 |
| 210722A | 50.20 | 1.145 |
| 210702A | 138.2 | 1.1757 |
| 210619B | 60.90 | 1.937 |
| 210610B | 69.38 | 1.13 |
| 210610A | 13.62 | 3.54 |
| 210517A | 3.06 | 2.486 |
| 210504A | 135.06 | 2.077 |
| 210420B | 158.8 | 1.400 |
| 210411C | 12.80 | 2.826 |
| 210321A | 8.21 | 1.487 |
| 210222B | 12.82 | 2.198 |
| 210210A | 6.60 | 0.715 |
| 201221A | 44.5 | 5.70 |
| 201216C | 48.0 | 1.10 |
| 201104B | 8.66 | 1.954 |
| 201024A | 5.00 | 0.999 |
| 201021C | 24.38 | 1.070 |
| 201020A | 14.17 | 2.903 |
| 201015A | 9.78 | 0.423 |

| Continued | | |
|---|---|---|
| 201014A | 36.2 | 4.56 |
| 200829A | 13.04 | 1.25 |
| 200205B | 458.0 | 1.465 |
| 191221B | 48.00 | 1.148 |
| 191019A | 64.35 | 0.248 |
| 191011A | 7.37 | 1.722 |
| 191004B | 37.7 | 3.503 |
| 190829A | 58.2 | 0.0785 |
| 190719C | 185.7 | 2.469 |
| 190324A | 28.4 | 1.1715 |
| 190114C | 361.5 | 0.42 |
| 190114A | 66.6 | 3.3765 |
| 190106A | 76.8 | 1.86 |
| 181110A | 138.4 | 1.505 |
| 181020A | 238.0 | 2.938 |
| 181010A | 16.4 | 1.39 |
| 180728A | 8.68 | 0.117 |
| 180624A | 486.4 | 2.855 |
| 180620B | 198.8 | 1.1175 |
| 180510B | 134.3 | 1.305 |
| 180404A | 35.2 | 1.000 |
| 180329B | 210.0 | 1.998 |
| 180325A | 94.1 | 2.25 |
| 180314A | 51.2 | 1.445 |
| 180205A | 15.5 | 1.409 |
| 180115A | 40.9 | 2.487 |
| 171222A | 174.8 | 2.409 |
| 171205A | 189.4 | 0.0368 |
| 171020A | 41.9 | 1.87 |
| 170903A | 29.2 | 0.886 |
| 170705A | 217.3 | 2.010 |
| 170604A | 26.70 | 1.329 |
| 170531B | 164.13 | 2.366 |
| 170519A | 216.4 | 0.818 |
| 170405A | 164.7 | 3.510 |
| Continued | | |
| 170202A | 46.2 | 3.645 |
| 170113A | 20.66 | 1.968 |
| 161219B | 6.94 | 0.1475 |
| 161129A | 35.53 | 0.645 |
| 161117A | 125.7 | 1.549 |
| 161108A | 105.1 | 1.159 |
| 161017A | 216.3 | 2.0127 |
| 161014A | 18.3 | 2.823 |
| 160804A | 144.2 | 0.736 |
| 160425A | 304.58 | 0.555 |
| 160410A | 8.2 | 1.717 |
| 160327A | 28 | 4.99 |
| 160314A | 8.73 | 0.726 |
| 160227A | 316.5 | 2.38 |
| 160203A | 20.2 | 3.52 |
| 160131A | 325 | 0.97 |
| 160121A | 12.0 | 1.960 |
| 151215A | 17.8 | 2.59 |
| 151112A | 19.32 | 4.1 |
| 151111A | 76.93 | 3.5 |
| 151031A | 5.00 | 1.167 |
| 151029A | 8.95 | 1.423 |
| 151027B | 80.00 | 4.063 |
| 151027A | 129.69 | 0.81 |
| 151021A | 110.2 | 2.330 |
| 150915A | 164.7 | 1.968 |
| 150910A | 112.2 | 1.359 |
| 150821A | 172.1 | 0.755 |
| 150818A | 123.3 | 0.282 |
| 150727A | 88 | 0.313 |
| 150413A | 263.6 | 3.2 |
| 150403A | 40.90 | 2.06 |
| 150323A | 149.6 | 0.593 |
| 150314A | 14.79 | 1.758 |
| 150301B | 12.44 | 1.5169 |

**Continued**

| | | |
|---|---|---|
| 150206A | 83.2 | 2.087 |
| 141225A | 40.24 | 0.915 |
| 141221A | 36.9 | 1.452 |
| 141220A | 7.21 | 1.3195 |
| 141121A | 549.9 | 1.47 |
| 141109A | 200 | 2.993 |
| 141026A | 146 | 3.35 |
| 141004A | 3.92 | 0.573 |
| 140907A | 79.2 | 1.21 |
| 140710A | 3.52 | 0.558 |
| 140703A | 67.1 | 3.14 |
| 140629A | 42.0 | 2.275 |
| 140614A | 720 | 4.233 |
| 140518A | 60.5 | 4.707 |
| 140515A | 23.4 | 6.32 |
| 140512A | 154.8 | 0.725 |
| 140506A | 111.1 | 0.889 |
| 140430A | 173.6 | 1.60 |
| 140428A | 17.42 | 4.7 |
| 140423A | 134 | 3.26 |
| 140419A | 94.7 | 3.956 |
| 140318A | 8.43 | 1.02 |
| 140311A | 71.4 | 4.95 |
| 140304A | 15.6 | 5.283 |
| 140301A | 31.0 | 1.416 |
| 140213A | 60.0 | 1.2076 |
| 140206A | 93.6 | 2.73 |
| 140114A | 139.7 | 3.0 |
| 131227A | 18.0 | 5.3 |
| 131117A | 11.00 | 4.042 |
| 131105A | 112.3 | 1.686 |
| 131103A | 17.3 | 0.599 |
| 131030A | 41.1 | 1.293 |
| 130907A | >360 | 1.238 |
| 130831A | 32.5 | 0.4791 |
| **Continued** | | |
| 130701A | 4.38 | 1.155 |
| 130612A | 4.0 | 2.006 |
| 130610A | 46.4 | 2.092 |
| 130606A | 276.58 | 5.913 |
| 130604A | 37.7 | 1.06 |
| 130514A | 204 | 3.6 |
| 130511A | 5.43 | 1.3033 |
| 130505A | 88 | 2.27 |
| 130427B | 27.0 | 2.78 |
| 130427A | 162.83 | 0.34 |
| 130420A | 123.5 | 1.297 |
| 130418A | >300 | 1.218 |
| 130408A | 28 | 3.758 |
| 130215A | 65.7 | 0.597 |
| 130131B | 4.30 | 2.539 |
| 121211A | 182 | 1.023 |
| 121201A | 85 | 3.385 |
| 121128A | 23.3 | 2.20 |
| 121027A | 62.6 | 1.773 |
| 121024A | 69 | 2.298 |
| 120922A | 173 | 3.1 |
| 120907A | 16.9 | 0.970 |
| 120815A | 9.7 | 2.358 |
| 120811C | 26.8 | 2.671 |
| 120805A | 48.00 | 3.1 |
| 120802A | 50 | 3.796 |
| 120729A | 71.5 | 0.80 |
| 120724A | 72.8 | 1.48 |
| 120722A | 42.4 | 0.9586 |
| 120714B | 159 | 0.3984 |
| 120712A | 14.7 | 4.1745 |
| 120521C | 26.7 | 6.0 |
| 120422A | 5.35 | 0.28 |
| 120404A | 38.7 | 2.876 |
| 120327A | 62.9 | 2.81 |

A. M. Hasan, W. J. Azzam

**Continued**

| | | |
|---|---|---|
| 120326A | 69.6 | 1.798 |
| 120119A | 253.8 | 1.728 |
| 120118B | 23.26 | 2.943 |
| 111229A | 25.4 | 1.3805 |
| 111228A | 101.20 | 0.714 |
| 111225A | 106.8 | 0.297 |
| 111123A | 290.0 | 3.1516 |
| 111107A | 26.6 | 2.893 |
| 111008A | 63.46 | 4.9898 |
| 110818A | 103 | 3.36 |
| 110808A | 48 | 1.348 |
| 110801A | 385 | 1.858 |
| 110731A | 38.8 | 2.83 |
| 110715A | 13.0 | 0.82 |
| 110503A | 10.0 | 1.613 |
| 110422A | 25.9 | 1.77 |
| 110213A | 48.0 | 1.46 |
| 110205A | 257 | 2.22 |
| 110128A | 30.7 | 2.339 |
| 101225A | 1088.0 | 0.847 |
| 101219B | 34 | 0.5519 |
| 100906A | 114.4 | 1.727 |
| 100902A | 428.8 | 4.5 |
| 100901A | 439 | 1.408 |
| 100816A | 2.9 | 0.8034 |
| 100814A | 174.5 | 1.44 |
| 100728B | 12.1 | 2.106 |
| 100728A | 198.5 | 1.567 |
| 100704A | 197.5 | 3.6 |
| 100621A | 63.6 | 0.542 |
| 100615A | 39 | 1.398 |
| 100513A | 84 | 4.772 |
| 100425A | 37.0 | 1.755 |
| 100424A | 104 | 2.465 |
| 100418A | 7.0 | 0.6235 |

Continued

| | | |
|---|---|---|
| 100413A | 191 | 3.9 |
| 100316B | 3.8 | 1.180 |
| 100302A | 17.9 | 4.813 |
| 100219A | 18.8 | 4.5 |
| 091208B | 14.9 | 1.063 |
| 091127 | 7.1 | 0.490 |
| 091109A | 48 | 3.076 |
| 091029 | 39.2 | 2.752 |
| 091024 | 109.8 | 1.092 |
| 091020 | 34.6 | 1.71 |
| 091018 | 4.4 | 0.971 |
| 090927 | 2.2 | 1.37 |
| 090926B | 109.7 | 1.24 |
| 090812 | 66.7 | 2.452 |
| 090809 | 5.4 | 2.737 |
| 090726 | 67.0 | 2.71 |
| 090715B | 266 | 3.00 |
| 090618 | 113.2 | 0.54 |
| 090530 | 48 | 1.266 |
| 090529 | >100 | 2.625 |
| 090519 | 64 | 3.9 |
| 090516A | 210 | 4.109 |
| 090424 | 48 | 0.544 |
| 090423 | 10.3 | 8.0 |
| 090418A | 56 | 1.608 |
| 090407 | 310 | 1.4485 |
| 090313 | 79 | 3.375 |
| 090205 | 8.8 | 4.7 |
| 090113 | 9.1 | 1.7493 |
| 090102 | 27.0 | 1.547 |
| 081222 | 24 | 2.77 |
| 081221 | 34 | 2.26 |
| 081203A | 294 | 2.1 |
| 081121 | 14 | 2.512 |
| 081118 | 67 | 2.58 |

Continued

| | | |
|---|---|---|
| 081029 | 270 | 3.8479 |
| 081028A | 260 | 3.038 |
| 081008 | 185.5 | 1.9685 |
| 081007 | 10.0 | 0.5295 |
| 080928 | 280 | 1.692 |
| 080916A | 60 | 0.689 |
| 080913 | 8 | 6.44 |
| 080906 | 147 | 2.0 |
| 080905B | 128 | 2.374 |
| 080810 | 106 | 3.35 |
| 080805 | 78 | 1.505 |
| 080804 | 34 | 2.2045 |
| 080721 | 16.2 | 2.602 |
| 080710 | 120 | 0.845 |
| 080707 | 27.1 | 1.23 |
| 080607 | 79 | 3.036 |
| 080605 | 20 | 1.6398 |
| 080604 | 82 | 1.416 |
| 080603B | 60 | 2.69 |
| 080520 | 2.8 | 1.545 |
| 080516 | 5.8 | 3.2 |
| 080430 | 16.2 | 0.767 |
| 080413B | 8.0 | 1.10 |
| 080413A | 46 | 2.433 |
| 080411 | 56 | 1.03 |
| 080330 | 61 | 1.51 |
| 080319C | 34 | 1.95 |
| 080319B | >50 | 0.937 |
| 080310 | 365 | 2.4266 |
| 080210 | 45 | 2.641 |
| 080207 | 340 | 2.0858 |
| 071122 | 68.7 | 1.14 |
| 071117 | 6.6 | 1.331 |
| 071112C | 15 | 0.8230 |
| 071031 | 180 | 2.692 |

| | Continued | |
|---|---|---|
| 071028B | 55 | 0.94 |
| 071021 | 225 | 2.4520 |
| 071020 | 4.2 | 2.142 |
| 071010B | >35.7 | 0.947 |
| 071010A | 6 | 0.98 |
| 071003 | 150 | 1.100 |
| 070810A | 11.0 | 2.17 |
| 070802 | 16.4 | 2.45 |
| 070714B | 64 | 0.92 |
| 070612A | 368.800 | 0.617 |
| 070611 | 12.200 | 2.04 |
| 070529 | 109.200 | 2.4996 |
| 070521 | 37.900 | 0.553 |
| 070508 | 20.900 | 0.82 |
| 070506 | 4.300 | 2.31 |
| 070419A | 115.600 | 0.97 |
| 070411 | 121.500 | 2.954 |
| 070318 | 74.600 | 0.836 |
| 070306 | 209.500 | 1.497 |
| 070208 | 47.700 | 1.165 |
| 070129 | 460.600 | 2.3384 |
| 070110 | 88.400 | 2.352 |
| 070103 | 18.600 | 2.6208 |
| 061222B | 40.000 | 3.355 |
| 061222A | 71.400 | 2.088 |
| 061210 | 85.300 | 0.41 |
| 061121 | 81.300 | 1.314 |
| 061110B | 134.000 | 3.44 |
| 061110A | 40.700 | 0.758 |
| 061021 | 46.200 | 0.3463 |
| 061007 | 75.300 | 1.261 |
| 060927 | 22.500 | 5.6 |
| 060926 | 8.000 | 3.208 |
| 060912A | 5.000 | 0.937 |
| 060908 | 19.300 | 1.8836 |

| Continued | | |
|---|---|---|
| 060906 | 43.500 | 3.685 |
| 060904B | 171.500 | 0.703 |
| 060814 | 145.300 | 0.84 |
| 060729 | 115.300 | 0.54 |
| 060719 | 66.900 | 1.5320 |
| 060714 | 115.000 | 2.71 |
| 060707 | 66.200 | 3.43 |
| 060614 | 108.700 | 0.125 |
| 060607A | 102.200 | 3.082 |
| 060605 | 79.100 | 3.78 |
| 060604 | 95.000 | 2.1357 |
| 060602A | 75.000 | 0.787 |
| 060526 | 298.200 | 3.21 |
| 060522 | 71.100 | 5.11 |
| 060512 | 8.500 | 0.4428 |
| 060510B | 275.200 | 4.9 |
| 060505 | ~4 | 0.089 |
| 060502A | 28.400 | 1.51 |
| 060418 | 103.100 | 1.490 |
| 060223A | 11.300 | 4.41 |
| 060218 | ~2100 | 0.0331 |
| 060210 | 255.000 | 3.91 |
| 060206 | 7.600 | 4.045 |
| 060124 | ~750 | 2.30 |
| 060123 | 900 | 1.099 |
| 060115 | 139.600 | 3.53 |
| 060108 | 14.300 | 2.03 |
| 051117B | 9.000 | 0.481 |
| 051111 | 46.100 | 1.549 |
| 051109B | 14.300 | 0.080 |
| 051109A | 37.200 | 2.346 |
| 051016B | 4.000 | 0.9364 |
| 051006 | 34.800 | 1.059 |
| 051001 | 189.100 | 2.4296 |
| 050922C | 4.500 | 2.198 |
| Continued | | |
| 050915A | 52.000 | 2.5273 |
| 050908 | 19.400 | 3.350 |
| 050904 | 174.200 | 6.10 |
| 050826 | 35.500 | 0.297 |
| 050824 | 22.600 | 0.83 |
| 050820A | 26 | 2.612 |
| 050819 | 37.700 | 2.5043 |
| 050814 | 150.900 | 5.3 |
| 050803 | 87.900 | 0.422 |
| 050802 | 19.000 | 1.71 |
| 050730 | 156.500 | 3.96855 |
| 050724 | 96.000 | 0.257 |
| 050525A | 8.800 | 0.606 |
| 050505 | 58.900 | 4.27 |
| 050416A | 2.500 | 0.6535 |
| 050406 | 5.400 | 2.44 |
| 050401 | 33.300 | 2.9 |
| 050319 | 152.500 | 3.24 |
| 050318 | 32 | 1.44 |
| 050315 | 95.600 | 1.949 |
| 050223 | 22.500 | 0.5915 |
| 050126 | 24.800 | 1.29 |

## Appendix B. MCMC Contour Plots

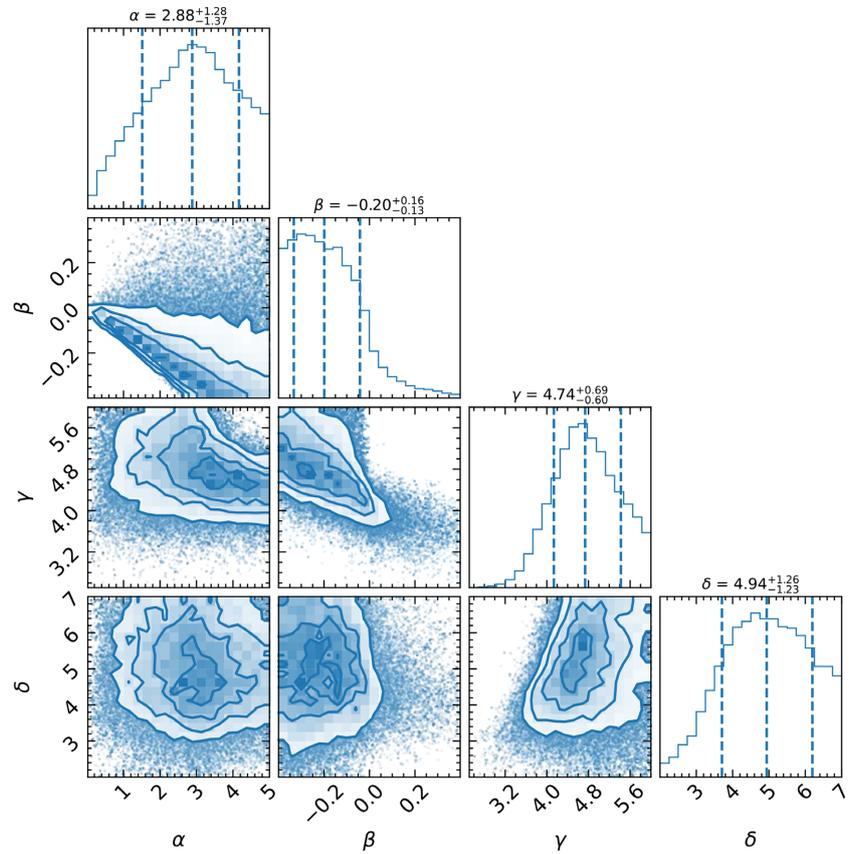

**Figure B1.** Shows the corner plot of the GRB-DD case.

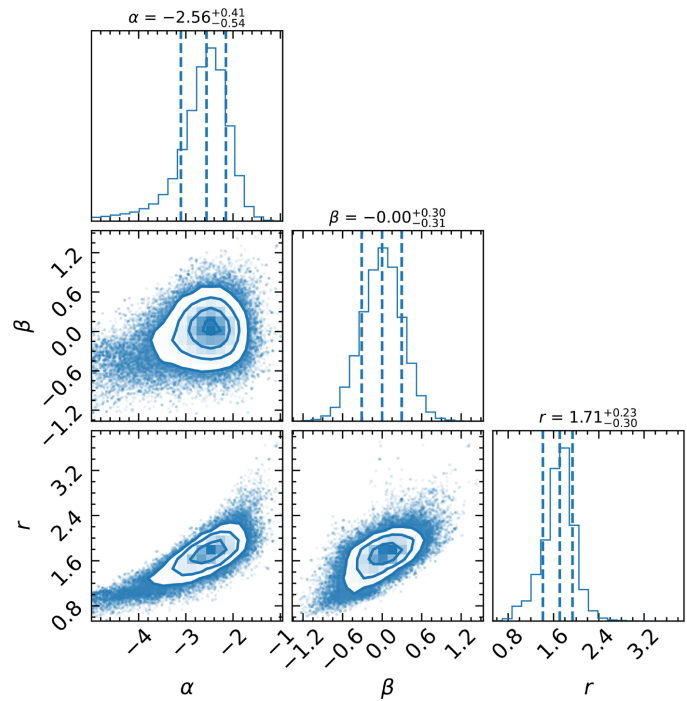

**Figure B2.** Shows the corner plot of the broken power law case.

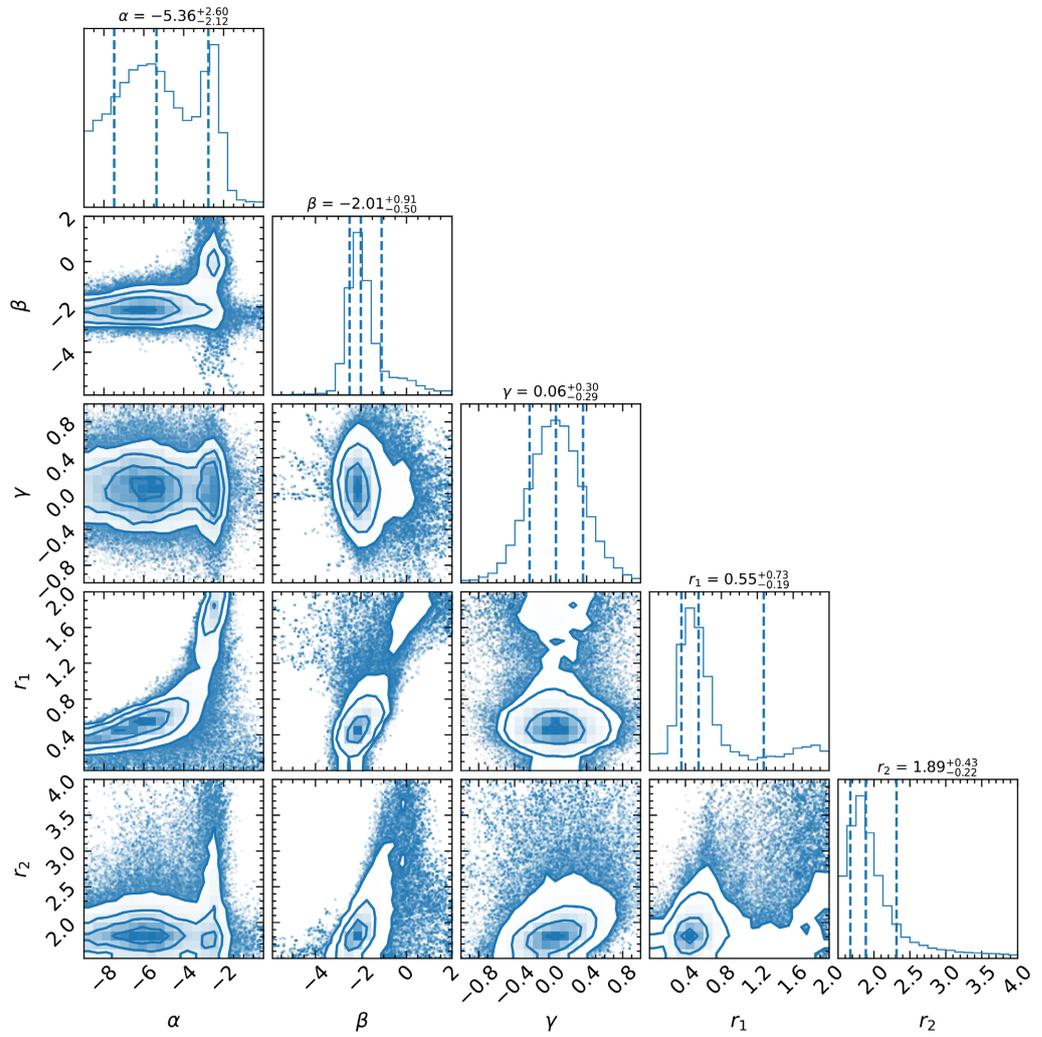

Figure B3. Shows the corner plot of the triple power law case.

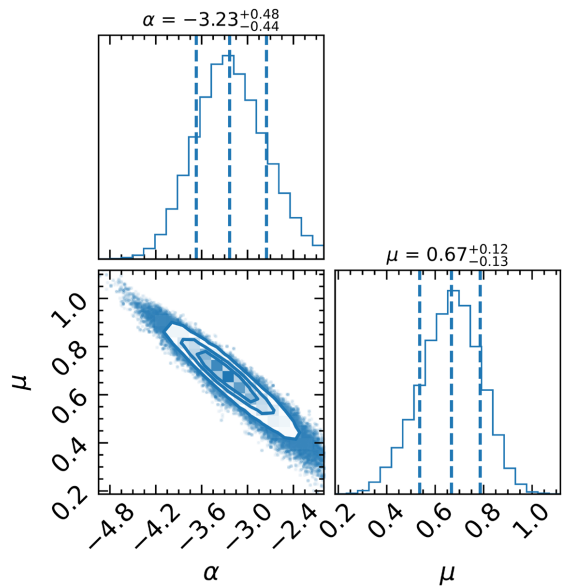

Figure B4. Shows the corner plot of the exponential-power law case.